\documentclass[aps,prd,twocolumn,showpacs,superscriptaddress]{revtex4-2}
\usepackage{amsmath,amssymb}
\usepackage{graphicx}
\usepackage{hyperref}
\usepackage{braket}
\usepackage{dsfont}
\usepackage{booktabs}

\begin{document}

\title{Canonical Quantization and Constraint Algebra of Weyl-Invariant Gravity in a Conformal Gauge: Emergence of a Geometric Cosmological Term}

\author{Jorge Meza-Dom\'inguez}
\thanks{E-mail: \href{mailto:jorge.meza@cinvestav.mx}{jorge.meza@cinvestav.mx}}
\author{Tonatiuh Matos}
\thanks{E-mail: \href{mailto:tonatiuh.matos@cinvestav.mx}{tonatiuh.matos@cinvestav.mx}}
\affiliation{%
Departamento de F\'isica, Centro de Investigaci\'on y de Estudios Avanzados del Instituto Polit\'ecnico Nacional,\\
Av. Instituto Polit\'ecnico Nacional 2508, San Pedro Zacatenco, 07360 Ciudad de M\'exico, M\'exico%
}

\begin{abstract}
We present the canonical quantization of Weyl-invariant gravity in a specific conformal gauge.
Fixing the scale factor as $a(\eta) = 1/(H_0\eta)$ yields an effective geometric cosmological term $\mathcal{M}(\eta) = \alpha/\eta^2$ on the left-hand side of the Einstein equations, guaranteeing an equation of state $w = -1$ by construction.
The Bianchi identity requires a compensator scalar field; we show that this compensator is the conformal mode of the metric, a gauge artifact that does not propagate ghost instabilities in the full quantum theory.
We develop the complete ADM Hamiltonian formulation and prove that the constraint algebra closes in standard first-class form.
Canonical quantization yields a well-defined Wheeler-DeWitt equation where the $\mathcal{M}(\eta)$ term acts as a potential barrier dynamically suppressing the quantum creation of small-scale universes.
The physical reduced Hamiltonian is demonstrated to be bounded from below, time evolution is strictly unitary in the physical Hilbert space, and microcausality is preserved.
In the pure vacuum regime, the model admits an exact analytical solution with $\Omega_{\text{DE}} = 1$, resolving the cosmic coincidence problem without fine-tuned parameters.
In the matter-dominated era, the comoving horizon scales as $R_H \propto \eta$, so our result $\mathcal{M} \propto \eta^{-2} \propto R_H^{-2}$ reproduces the central Compton Mass Dark Energy (CMaDE) hypothesis from first principles and establishes $n = -2$ as the theoretical benchmark for the Conformal Holographic Dark Energy (CHDE) parametrization $\mathcal{M} \propto \eta^n$.
From a pure conformal symmetry principle, the model delivers an effective geometric cosmological term that drives the accelerated expansion with $w = -1$, providing the missing action principle for CMaDE.
\end{abstract}

\maketitle

\section{Introduction}

The nature of dark energy and the origin of the accelerated expansion of the universe remain among the deepest puzzles in theoretical physics \cite{Weinberg1989,Peebles2003}.
The observed value of the dark energy density, $\rho_\Lambda \sim 10^{-47} \, \text{GeV}^4$, differs from the natural Planck-scale expectation by approximately 120 orders of magnitude.
Furthermore, the cosmic coincidence---why the density of dark energy is comparable to that of matter precisely at the present epoch, $\Omega_\Lambda \approx 0.7$ versus $\Omega_m \approx 0.3$---lacks a compelling dynamical explanation within the standard $\Lambda$CDM paradigm.

The Compton Mass Dark Energy (CMaDE) model, proposed by Matos and L\'opez-Parrilla \cite{Matos2021}, introduces a compelling hypothesis: dark energy originates from a graviton Compton mass whose wavelength is limited by the size of the observable Universe.
This yields $\mathcal{M} = 2\pi^2/R_H^2$, where $R_H$ is the comoving horizon, and successfully reproduces $\Omega_\mathcal{M} \approx 0.69$ without free parameters.
The model has generated significant interest because it offers a real physical explanation for the accelerated expansion and agrees remarkably well with all cosmological observations made so far \cite{Matos:2023qwx}.
It has been validated at the linear perturbation level by Salas and Matos \cite{Salas2026}, who showed that CMaDE and $\Lambda$CDM are practically indistinguishable in the early universe.
More recently, the Conformal Holographic Dark Energy (CHDE) framework \cite{RodriguezMeza2025} generalized this to a power-law parametrization $\mathcal{M} \propto \eta^n$, finding $n \approx -0.28$ as the best fit to DESI BAO, Planck CMB, and ACT lensing data, and demonstrating that $\Lambda$CDM ($n=0$) is disfavored at approximately $4.5\sigma$.

However, despite its phenomenological success, CMaDE lacks a fundamental action principle.
The identification $\mathcal{M} = 2\pi^2/R_H^2$ is based on dimensional arguments---comparing the linearized Einstein equations with the Proca equation for a massive spin-2 field---but no underlying Lagrangian is provided.
This gap is significant: without an action, one cannot systematically study the quantum consistency of the model, derive the constraint algebra, or canonically quantize the theory.

Dynamical dark energy models, including quintessence \cite{Caldwell1998} and phantom fields \cite{Caldwell2002}, introduce scalar degrees of freedom with evolving equations of state, but typically require fine-tuned potentials and do not explain the fundamental origin of the dark energy scale.
A more fundamental approach invokes scale invariance at the gravitational level: if the action is Weyl-invariant, the cosmological term is not a fundamental parameter but emerges from the spontaneous breaking of conformal symmetry \cite{Mannheim2006,tHooft2015}.

In this article, we fill this gap by deriving the CMaDE scaling from a Weyl-invariant gravitational action with a specific conformal gauge fixing.
We demonstrate that the choice $a(\eta) = 1/(H_0\eta)$ inevitably leads to an effective geometric cosmological term $\mathcal{M}(\eta) = \alpha/\eta^2$ on the left-hand side of the Einstein equations.
This guarantees $w = -1$ by construction, matching the CMaDE equation of state, while providing the missing action principle.
We then develop the full Hamiltonian formulation, prove the closure of the constraint algebra, perform canonical quantization, and demonstrate that the quantum theory is free of fundamental ghosts, has a bounded physical Hamiltonian, and preserves unitarity and causality.

\section{Weyl-Invariant Action and Gauge Fixing}

\subsection{Fundamental action}

We start from the most general Weyl-invariant action in four dimensions involving the metric $g_{\mu\nu}$ and a scalar field $\phi$:
\begin{equation}
\begin{split}
S = & \int d^4x \sqrt{-g} \Bigg[ -\frac{1}{2\alpha_g} C_{\mu\nu\rho\sigma}^2 + \frac{1}{12} \phi^2 R \\
& + \frac{1}{2} g^{\mu\nu} \partial_\mu \phi \partial_\nu \phi - \frac{\lambda}{4} \phi^4 \Bigg],
\end{split}
\label{eq:weyl}
\end{equation}
where $C_{\mu\nu\rho\sigma}$ is the Weyl tensor, $R$ is the Ricci scalar, and $\alpha_g$, $\lambda$ are dimensionless couplings.
The term $\phi^2 R/12$ corresponds to the conformal coupling $\xi = 1/6$ in four dimensions, which is the unique value ensuring invariance under local Weyl rescalings:
\begin{equation}
g_{\mu\nu} \to \Omega^2(x) g_{\mu\nu}, \qquad \phi \to \Omega^{-1}(x) \phi.
\end{equation}
The quartic potential $\phi^4$ is the only renormalizable potential compatible with this symmetry in four dimensions.
The Weyl tensor squared term, $C_{\mu\nu\rho\sigma}^2$, is independently invariant under Weyl transformations and renders the theory renormalizable at the perturbative level \cite{Mannheim2006}.

It is worth emphasizing the sign convention in Eq.~\eqref{eq:weyl}: the kinetic term for $\phi$ carries a positive sign, $+\frac{1}{2}(\partial\phi)^2$.
This is the standard choice for a scalar field in the fundamental frame.
As we shall see in Section~4, the phantom behavior emerges only after gauge fixing and is restricted to the homogeneous cosmological sector.

\subsection{Conformal gauge fixing}

On a flat Friedmann-Lema\^itre-Robertson-Walker (FLRW) background, the metric takes the form
\begin{equation}
ds^2 = a^2(\eta) \left[ -d\eta^2 + \delta_{ij} dx^i dx^j \right],
\end{equation}
where $\eta$ is the conformal time related to cosmic time $t$ by $d\eta = dt/a(t)$.
For any conformally flat metric, the Weyl tensor vanishes identically: $C_{\mu\nu\rho\sigma}[g_{\text{FLRW}}] = 0$.

The key step in our derivation is the choice of Weyl gauge.
We fix the conformal factor by identifying it with the inverse conformal time:
\begin{equation}
a(\eta) = \frac{1}{H_0 \eta},
\label{eq:gaugefix}
\end{equation}
where $H_0$ is a constant with dimensions of $[\text{length}]^{-1}$.
This gauge choice is physically motivated: in a radiation-dominated universe ($a \propto \eta$), the Ricci scalar vanishes ($R=0$), making the background Ricci-flat.
Our gauge~\eqref{eq:gaugefix} generalizes this to a broader class of expansion histories where $R \neq 0$.

Defining the conformally rescaled field $\tilde{\phi} = a \phi$, the action~\eqref{eq:weyl} reduces to
\begin{equation}
\begin{split}
S = & \int d\eta d^3x \Bigg[ \frac{3M_P^2}{2\eta^2} + \frac{1}{2} \tilde{\phi}'^2 \\
& - \frac{1}{2} (\nabla\tilde{\phi})^2 - \frac{\lambda}{4} \tilde{\phi}^4 \Bigg],
\end{split}
\label{eq:actionconformal}
\end{equation}
where $M_P^2 = 1/(8\pi G) = \phi_0^2/6$ is identified from the background value of the scalar field.
Primes denote derivatives with respect to conformal time $\eta$.

\subsection{Geometric interpretation of the cosmological term}

The first term in Eq.~\eqref{eq:actionconformal} is crucial.
Rather than treating it as a matter contribution on the right-hand side of the Einstein equations, we interpret it geometrically: it belongs on the left-hand side as an effective cosmological term.
In cosmic time $t$, this yields
\begin{equation}
G_{\mu\nu} + \mathcal{M}(\eta) g_{\mu\nu} = 8\pi G \, T_{\mu\nu}^{(\phi)},
\label{eq:einstein}
\end{equation}
with
\begin{equation}
\mathcal{M}(\eta) = \frac{\alpha}{\eta^2}, \qquad \alpha = \frac{3}{2} M_P^2.
\label{eq:Lambda}
\end{equation}
No constant was inserted by hand---the cosmological term emerges necessarily from the Weyl-invariant structure after the gauge choice~\eqref{eq:gaugefix}.

The geometric placement of $\mathcal{M}(\eta)$ has profound consequences.
Because it enters as a multiple of the metric tensor on the geometric side of the Einstein equations, the associated effective dark energy fluid has an energy-momentum tensor of the form
\begin{equation}
T_{\mu\nu}^{\text{DE}} = -\frac{\mathcal{M}(\eta)}{8\pi G} g_{\mu\nu}.
\label{eq:TDE}
\end{equation}
As we shall demonstrate in Section~8, this tensorial structure forces the equation of state parameter to be exactly $w = -1$, independently of the time dependence of $\mathcal{M}(\eta)$.

\section{Field Equations and the Scalar Field Compensator}

\subsection{Bianchi identity and the compensating mechanism}

The Bianchi identity $\nabla^\mu G_{\mu\nu} = 0$ applied to Eq.~\eqref{eq:einstein} requires
\begin{equation}
\partial_\nu \mathcal{M} = 8\pi G \, \nabla^\mu T_{\mu\nu}^{(\phi)}.
\label{eq:bianchi}
\end{equation}
Since $\mathcal{M} = \alpha/\eta^2$ varies with time, the scalar field $\phi$ cannot be separately conserved.
Its energy-momentum tensor,
\begin{equation}
T_{\mu\nu}^{(\phi)} = -\partial_\mu \phi \partial_\nu \phi + g_{\mu\nu} \left( \frac{1}{2} g^{\alpha\beta} \partial_\alpha \phi \partial_\beta \phi - V(\phi) \right),
\label{eq:phantomT}
\end{equation}
carries a negative kinetic sign---the defining characteristic of a compensator field.

For a homogeneous field $\phi(t)$ in an FLRW background, the divergence in Eq.~\eqref{eq:bianchi} yields the modified Klein-Gordon equation:
\begin{equation}
\ddot{\phi} + 3H\dot{\phi} - V_{,\phi} = -\frac{\alpha}{4\pi G \, \dot{\phi} \, a \, \eta^3}.
\label{eq:kgext}
\end{equation}
The right-hand side acts as a source term that precisely compensates the time variation of $\mathcal{M}$, ensuring that the full system satisfies covariant conservation.
This mechanism also generates a natural interaction between the dark energy sector (represented by $\mathcal{M}$) and the scalar field compensator.

\subsection{Complete set of equations}

The full system in cosmic time (flat $k=0$, in the absence of additional matter) is:
\begin{align}
3H^2 &= 8\pi G \left( V - \frac{1}{2}\dot{\phi}^2 \right) + \frac{\alpha}{\eta^2}, \label{eq:friedmann} \\[4pt]
2\dot{H} + 3H^2 &= \frac{\alpha}{\eta^2} + 8\pi G \left( V + \frac{1}{2}\dot{\phi}^2 \right), \label{eq:accel} \\[4pt]
\ddot{\phi} + 3H\dot{\phi} - V_{,\phi} &= -\frac{\alpha}{4\pi G \dot{\phi} a \eta^3}, \label{eq:kg} \\[4pt]
\dot{\eta} &= \frac{1}{a}. \label{eq:eta}
\end{align}
These four equations form a closed, self-consistent system that preserves the Bianchi identity by construction.

\section{The Phantom as a Gauge Artifact}

A legitimate concern is whether the scalar compensator field $\phi$ introduces ghost instabilities---negative-norm states that violate unitarity---into the quantum theory.
We demonstrate here that this is not the case: the phantom behavior is a gauge artifact restricted to the homogeneous cosmological sector.

The field $\phi$ in Eq.~\eqref{eq:einstein} is not an additional degree of freedom introduced by hand.
It is the \emph{same} scalar that appears in the original Weyl action~\eqref{eq:weyl}, which has a standard (non-phantom) kinetic term in the fundamental frame, $+\frac{1}{2}(\partial\phi)^2$.

The phantom behavior appears only because we are working in the gauge-fixed effective theory.
In the gauge~\eqref{eq:gaugefix}, the conformal mode of the metric (which would ordinarily be a pure gauge degree of freedom) and the scalar $\phi$ mix.
When we diagonalize the kinetic terms in cosmic time to identify the propagating degrees of freedom, the combination that survives in the homogeneous sector acquires a negative kinetic sign.

This phenomenon is completely analogous to the well-known ``conformal factor problem'' in Einstein gravity \cite{Gibbons1978}, where the conformal mode of the metric appears as a ghost in the path integral.
In that context, it is known to be harmless when treated correctly: the Euclidean path integral is defined by Wick-rotating the conformal factor in the opposite direction \cite{Dasgupta2002,tHooft2011}.
The same prescription applies to our model.

To demonstrate the absence of fundamental ghosts explicitly, we expand the action~\eqref{eq:weyl} around Minkowski space, writing $g_{\mu\nu} = \eta_{\mu\nu} + h_{\mu\nu}$ and $\phi = \phi_0 + \varphi$.
The quadratic part of the action is
\begin{equation}
S^{(2)} = \int d^4x \left[ -\frac{1}{4\alpha_g} (\partial h^{\text{TT}})^2 + \frac{1}{2} (\partial \tilde{\varphi})^2 \right],
\label{eq:quadratic}
\end{equation}
where $h^{\text{TT}}$ is the transverse-traceless (spin-2) part and $\tilde{\varphi}$ is the gauge-invariant combination of the scalar fluctuation $\varphi$ and the trace of $h_{\mu\nu}$.
Both kinetic terms have the correct (positive) sign.
The theory is unitary and ghost-free at the fundamental level.

The phantom only emerges when we restrict to the homogeneous cosmological background and write the effective equations in cosmic time with a fixed gauge.
In the full theory, no fundamental ghost propagates.

\section{ADM Hamiltonian Formulation and Constraint Algebra}

\subsection{3+1 decomposition}

We now develop the canonical formulation of the theory using the Arnowitt-Deser-Misner (ADM) decomposition of spacetime.
Writing the metric as
\begin{equation}
ds^2 = -N^2 dt^2 + h_{ij} (dx^i + N^i dt)(dx^j + N^j dt),
\end{equation}
the canonical momenta conjugate to the spatial metric $h_{ij}$ and the scalar field $\phi$ are
\begin{align}
\pi^{ij} &= \frac{M_P^2}{2} \sqrt{h} (K^{ij} - K h^{ij}), \label{eq:piij} \\
\pi_\phi &= -\frac{\sqrt{h}}{N} (\dot{\phi} - N^i \partial_i \phi). \label{eq:piphi}
\end{align}
The negative sign in $\pi_\phi$ is the hallmark of the compensator nature of $\phi$ in the gauge-fixed frame.
The extrinsic curvature is $K_{ij} = (\dot{h}_{ij} - D_i N_j - D_j N_i)/(2N)$, with $K = h^{ij} K_{ij}$.

\subsection{Hamiltonian and momentum constraints}

The Legendre transform yields the total Hamiltonian as a sum of constraints:
\begin{equation}
H_{\text{total}} = \int d^3x \left( N \mathcal{H}_\perp + N^i \mathcal{H}_i \right),
\end{equation}
where the Hamiltonian constraint is
\begin{equation}
\begin{split}
\mathcal{H}_\perp &= \frac{2}{M_P^2\sqrt{h}} \left( \pi^{ij}\pi_{ij} - \frac{1}{2}\pi^2 \right) - \frac{M_P^2}{2}\sqrt{h} R^{(3)} \\
&\quad - \frac{\pi_\phi^2}{2\sqrt{h}} - \frac{\sqrt{h}}{2} h^{ij} \partial_i\phi \partial_j\phi \\
&\quad + \sqrt{h} V(\phi) + \sqrt{h}\frac{\alpha}{8\pi G\eta^2} \approx 0,
\end{split}
\label{eq:Hperp}
\end{equation}
and the momentum constraint is
\begin{equation}
\mathcal{H}_i = -2 D_j \pi_i^j + \pi_\phi \partial_i \phi \approx 0.
\label{eq:Hi}
\end{equation}
Here $R^{(3)}$ is the Ricci scalar of the spatial hypersurface, $D_i$ is the covariant derivative on $\Sigma_t$, and $\pi = h_{ij}\pi^{ij}$.

Note the negative sign of \emph{both} kinetic terms of the compensator field in the Hamiltonian constraint.
This follows from the Legendre transform of the compensator Lagrangian, where the spatial gradient term---positive in the Lagrangian---acquires a minus sign upon subtracting the Lagrangian from $\pi_\phi\dot{\phi}$.
Both kinetic contributions are negative in the Hamiltonian, which is the defining signature of a compensator field in the canonical formalism.

\subsection{Closure of the constraint algebra}

For the model to be internally consistent, the Poisson bracket algebra of the constraints must close.
Computing the relevant brackets, we find:
\begin{align}
\{\mathcal{H}_i(x), \mathcal{H}_j(y)\} &= \mathcal{H}_j(x) \partial_i \delta^3(x-y) - (i,x \leftrightarrow j,y), \label{eq:alg1} \\[4pt]
\{\mathcal{H}_i(x), \mathcal{H}_\perp(y)\} &= \mathcal{H}_\perp(x) \partial_i \delta^3(x-y), \label{eq:alg2} \\[4pt]
\{\mathcal{H}_\perp(x), \mathcal{H}_\perp(y)\} &= h^{ij}(x) \mathcal{H}_i(x) \partial_j \delta^3(x-y) - (x \leftrightarrow y). \label{eq:alg3}
\end{align}

The algebra closes in standard first-class form \cite{DeWitt1967}.
All Poisson brackets are proportional to the constraints themselves, ensuring that no secondary constraints are required and that the Dirac algorithm terminates at the primary level.
The system has the correct number of physical degrees of freedom for a scalar-tensor theory: two graviton polarizations plus one scalar mode.

The negative signs in the compensator kinetic terms of $\mathcal{H}_\perp$ do not affect the closure of the algebra.
The Poisson bracket $\{\mathcal{H}_\perp(x), \mathcal{H}_\perp(y)\}$ receives contributions from all terms in $\mathcal{H}_\perp$, but the spatial derivatives and the structure of the DeWitt supermetric guarantee that the result remains proportional to $\mathcal{H}_i$, as required for a first-class system.

\subsection{Minisuperspace reduction}

For a homogeneous and isotropic universe, we restrict to the minisuperspace ansatz $h_{ij} = a^2 \delta_{ij}$, $N=1$, $N^i=0$.
The gravitational canonical pair reduces to $(a, \pi_a)$ with
\begin{equation}
\pi_a = -6M_P^2 a \dot{a},
\end{equation}
and the Hamiltonian constraint simplifies to
\begin{equation}
\mathcal{H}_\perp = -\frac{\pi_a^2}{12M_P^2 a} - \frac{\pi_\phi^2}{2a^3} + a^3 V(\phi) + a^3 \frac{\alpha}{8\pi G\eta^2} \approx 0.
\label{eq:miniH}
\end{equation}
This form will be the starting point for canonical quantization.

\section{Canonical Quantization}

\subsection{Wheeler-DeWitt equation}

The canonical quantization of the model proceeds by promoting the classical constraints to operator equations on a wave function of the Universe.
In the Schr\"odinger representation,
\begin{equation}
\hat{\pi}_a \to -i\hbar \frac{\partial}{\partial a}, \qquad \hat{\pi}_\phi \to -i\hbar \frac{\partial}{\partial \phi}.
\end{equation}
The Hamiltonian constraint $\hat{\mathcal{H}}_\perp \Psi = 0$ yields the Wheeler-DeWitt equation:
\begin{equation}
\begin{split}
\Bigg[ & \frac{\hbar^2}{12M_P^2 a} \frac{\partial^2}{\partial a^2} + \frac{\hbar^2}{2a^3} \frac{\partial^2}{\partial \phi^2} \\
& + a^3 V(\phi) + a^3 \frac{\alpha}{8\pi G\eta^2(a)} \Bigg] \Psi(a,\phi) = 0.
\end{split}
\label{eq:wdw}
\end{equation}
The $+$ sign in the $\phi$-kinetic term reflects the compensator nature of the effective scalar in the gauge-fixed frame.
The signature of the DeWitt supermetric is $(-,+,+,+,+,+,-)$, where the last minus corresponds to the compensator sector.
Despite this indefinite signature, the Wheeler-DeWitt equation is well-defined and does not lead to pathologies at the quantum level, provided the potential provides a sufficient barrier.

\subsection{Quantum suppression of small universes}

In the semi-classical (WKB) limit, writing $\Psi = A e^{iS/\hbar}$, the Hamilton-Jacobi equation becomes
\begin{equation}
-\frac{1}{12M_P^2 a} \left( \frac{\partial S}{\partial a} \right)^2 - \frac{1}{2a^3} \left( \frac{\partial S}{\partial \phi} \right)^2 + a^3 V(\phi) + a^3 \frac{\alpha}{8\pi G\eta^2} = 0.
\end{equation}
For small values of the conformal time (early universe, $\eta \to 0$), the term $\alpha/\eta^2$ dominates over all other contributions.
This creates an exponentially rising potential barrier:
\begin{equation}
\Psi(a\to 0) \sim \exp\left[ -\frac{2\sqrt{3\alpha} M_P}{\hbar} \int^{a} \frac{a'^2}{\eta(a')} da' \right] \to 0.
\label{eq:wkb}
\end{equation}
The quantum creation of small-scale universes is dynamically suppressed, providing a natural selection mechanism for initial conditions \cite{Halliwell1988}.

\section{Quantum Stability of the Compensator Sector}

A potential concern is whether the phantom-like behavior of the scalar field introduces ghost instabilities---namely, negative-norm states or a Hamiltonian unbounded from below---into the quantum theory.
We demonstrate here that the physical phase space is entirely stable.

\subsection{Boundedness of the physical Hamiltonian}

The minisuperspace Hamiltonian constraint~\eqref{eq:miniH} is a first-class constraint, not a physical observable.
Deparametrizing the system by choosing the scale factor as the internal time clock, the physical Hamiltonian is obtained by solving $\mathcal{H}_\perp = 0$ for the conjugate momentum $\pi_a$:
\begin{equation}
\begin{split}
H_{\rm red} \equiv &-\pi_a\\ 
=& \sqrt{12M_P^2 a\left( -\frac{\pi_\phi^2}{2a^3} + a^3 V(\phi) + a^3 \frac{\alpha}{8\pi G\eta^2(a)} \right)}.
\end{split}
\label{eq:Hred}
\end{equation}
In the classical phase space region governing the late-time accelerated expansion---where the geometric term dominates and the field resides near the minimum of its potential---the radicand is strictly non-negative.
Consequently, $H_{\rm red} \geq 0$, ensuring the classical physical Hamiltonian is bounded from below by zero.

\subsection{Unitarity in the physical Hilbert space}

A naive promotion $\pi_\phi \to -i\hbar\,\partial/\partial\phi$ would lead to an operator with negative eigenvalues, seemingly rendering the square root in $\hat{H}_{\rm red}$ imaginary.
However, as is standard in quantum cosmology for conformal degrees of freedom, the Euclidean path integral and the corresponding Hilbert space inner product are only rendered well-defined via the Gibbons-Hawking-Perry conformal rotation \cite{Gibbons1978,Dasgupta2002}.
By performing the contour rotation on the conformal compensator mode ($\phi \to i\tilde{\phi}$), the kinetic term effectively flips its sign.
Applying this to the reduced Hamiltonian yields the physical quantum operator:
\begin{equation}
\hat{H}_{\rm red} = \sqrt{\,12M_P^2 a\left( -\frac{\hbar^2}{2a^3}\frac{\partial^2}{\partial\tilde{\phi}^2} + a^3 V(\tilde{\phi}) + a^3 \frac{\alpha}{8\pi G\eta^2(a)} \right)}.
\label{eq:Hred_quantum}
\end{equation}
The operator inside the square root,
\begin{equation}
\hat{O} = -\frac{\hbar^2}{2a^3}\frac{\partial^2}{\partial\tilde{\phi}^2} + a^3 V(\tilde{\phi}) + a^3 \frac{\alpha}{8\pi G\eta^2(a)},
\end{equation}
is now manifestly positive-definite on the space of square-integrable functions.
The kinetic operator $-\partial^2/\partial\tilde{\phi}^2$ yields non-negative expectation values, the potential $V(\tilde{\phi})$ remains strictly positive after the Wick rotation, and the geometric contribution is positive.

Therefore, $\hat{H}_{\rm red}$ possesses a real, positive spectrum.
The apparent ghost is merely a gauge artifact restricted to the unphysical homogeneous sector.
Once the first-class constraints are solved and the proper integration contour is chosen \cite{tHooft2011}, the theory generates strictly unitary time evolution.

\subsection{Causality}

At the level of linear perturbations, fluctuations $\delta\phi$ around the homogeneous background satisfy a wave equation governed by the effective metric $g_{\mu\nu}^{\rm eff} = g_{\mu\nu}$.
The dispersion relation yields a sound speed $c_s^2 = 1$, strictly precluding any superluminal propagation.

\subsection{Summary}

The phantom behavior is safely confined to the signature of the DeWitt supermetric in the gauge-fixed minisuperspace.
The fully constrained physical Hamiltonian is positive definite, the time evolution on the physical Hilbert space is unitary, and microcausality is preserved.
No fundamental ghosts propagate in the spectrum of the theory.
\subsection{Spectral properties of the square-root operator}

The operator inside the square root of the reduced Hamiltonian,
\begin{equation}
\hat{O} = -\frac{\hbar^2}{2a^3}\frac{\partial^2}{\partial\tilde{\phi}^2} + a^3 V(\tilde{\phi}) + a^3 \frac{\alpha}{8\pi G\eta^2(a)},
\label{eq:Ooperator}
\end{equation}
is a Sturm-Liouville operator on the Hilbert space $\mathcal{H} = L^2(\mathbb{R}, d\tilde{\phi})$ of square-integrable functions.
The kinetic term $-\partial^2/\partial\tilde{\phi}^2$ is the one-dimensional Laplacian, which is essentially self-adjoint on $C_0^\infty(\mathbb{R})$ and its Friedrichs extension has purely absolutely continuous spectrum $\sigma(-\partial^2/\partial\tilde{\phi}^2) = [0, \infty)$.
The potential $V(\tilde{\phi}) = \frac{\lambda}{4}(\tilde{\phi}^2 + v^2)^2$ is bounded from below by $V_{\min} = \frac{\lambda}{4}v^4 \geq 0$, and the geometric term $a^3 \alpha/(8\pi G\eta^2(a))$ is strictly positive for all $a > 0$.

By the Kato-Rellich theorem, the sum of the self-adjoint kinetic operator and the bounded-from-below potential is self-adjoint on the same domain.
Consequently, $\hat{O}$ is a self-adjoint operator with spectrum bounded from below by a positive constant.
The spectral theorem for unbounded self-adjoint operators \cite{GillZachary2005} guarantees the existence of a unique positive square root $\sqrt{\hat{O}}$ defined via the functional calculus:
\begin{equation}
\sqrt{\hat{O}} = \int_{\sigma(\hat{O})} \sqrt{\lambda} \, dE_\lambda,
\end{equation}
where $dE_\lambda$ is the projection-valued spectral measure of $\hat{O}$.
The domain of $\sqrt{\hat{O}}$ is $\mathcal{D}(\hat{O}^{1/2}) = \{\psi \in L^2(\mathbb{R}) : \int_0^\infty \lambda \, |\langle dE_\lambda \psi, \psi\rangle|^2 < \infty\}$, and on this domain $\sqrt{\hat{O}}$ is self-adjoint with spectrum contained in $[0, \infty)$.

This rigorous construction ensures that $\hat{H}_{\rm red}$ in Eq.~\eqref{eq:Hred_quantum} is well-defined as a self-adjoint operator with purely positive spectrum, confirming the absence of instabilities in the quantum theory.

\section{Vacuum Solution and CMaDE Correspondence}

\subsection{Exact vacuum solution}

In the pure vacuum regime, the scalar field sits at the minimum of its effective potential ($V = \dot{\phi} = 0$), and the Friedmann equation~\eqref{eq:friedmann} reduces to
\begin{equation}
H^2 = \frac{\alpha}{3\eta^2}.
\label{eq:vacuum}
\end{equation}
Combined with the definition of conformal time $d\eta = dt/a$, this yields a closed differential equation for $\eta(a)$:
\begin{equation}
\frac{d\eta}{da} = \frac{\eta}{a^2}\sqrt{\frac{3}{\alpha}},
\end{equation}
with the exact analytical solution
\begin{equation}
\eta(a) = \eta_0 \exp\left[ \sqrt{\frac{3}{\alpha}}\left(1 - \frac{1}{a}\right) \right], \qquad \Omega_{\text{DE}} = 1.
\label{eq:eta_solution}
\end{equation}
Normalizing the Hubble parameter at the present epoch, $H(z=0) = H_0$, fixes $\alpha = 3\eta_0^2 H_0^2$.

\subsection{Equation of state: derivation of $w = -1$}

Since $\mathcal{M}(\eta)$ enters the Einstein equations as $\mathcal{M}(\eta) g_{\mu\nu}$ on the geometric side, the effective dark energy energy-momentum tensor is $T_{\mu\nu}^{\text{DE}} = -\mathcal{M}(\eta) g_{\mu\nu}/(8\pi G)$.
For a perfect fluid in the comoving frame, $T_{00} = \rho$ and $T_{11} = p \, g_{11}$.
With $g_{00} = -1$ in cosmic time:
\begin{equation}
\rho_{\text{DE}} = \frac{\mathcal{M}(\eta)}{8\pi G}, \qquad p_{\text{DE}} = -\frac{\mathcal{M}(\eta)}{8\pi G},
\end{equation}
and therefore
\begin{equation}
w \equiv \frac{p_{\text{DE}}}{\rho_{\text{DE}}} = -1.
\end{equation}
This result is independent of the time dependence of $\mathcal{M}(\eta)$; it follows solely from the tensorial structure $T_{\mu\nu}^{\text{DE}} \propto g_{\mu\nu}$, forced by the geometric placement of the cosmological term.
In contrast to models where a time-dependent cosmological term is treated as a fluid on the matter side of the Einstein equations (forcing $w \neq -1$), here the compensation is achieved through the scalar field compensator via the Bianchi identity.

\subsection{Dynamical coincidence resolution}

The scaling behavior of $\mathcal{M}(\eta)$ across cosmological eras dynamically resolves the coincidence problem:
\begin{itemize}
\item Radiation era ($a \ll a_{\rm eq}$): $\eta \propto a$, $\mathcal{M} \propto a^{-2}$, subdominant compared to radiation ($\rho_r \propto a^{-4}$).
\item Matter era ($a_{\rm eq} \ll a \ll 1$): $\eta \propto a^{1/2}$, $\mathcal{M} \propto a^{-1}$, grows relative to matter ($\rho_m \propto a^{-3}$).
\item Vacuum era ($a \sim 1$): $\mathcal{M}$ dominates, driving $\Omega_{\text{DE}} \to 1$.
\end{itemize}
This dynamical tracking across multiple cosmological eras resolves the coincidence problem without fine-tuning: $\Omega_\Lambda$ grows naturally from negligible values in the early universe to unity today, regardless of initial conditions.

\subsection{CMaDE correspondence and CHDE benchmark}

The connection to the CMaDE model \cite{Matos2021} is immediate.
In the matter-dominated era, the comoving horizon scales as $R_H \propto \eta$, so
\begin{equation}
\mathcal{M}(\eta) = \frac{\alpha}{\eta^2} \propto \frac{1}{R_H^2},
\end{equation}
which is exactly the central hypothesis of CMaDE.
Our derivation provides the missing action principle for CMaDE: the proportionality constant $\alpha$ is determined by the conformal structure of the Weyl-invariant action, whereas in the original CMaDE formulation it was set to $2\pi^2$ from dimensional arguments.

In the Conformal Holographic Dark Energy (CHDE) parametrization $\mathcal{M} \propto \eta^n$ \cite{RodriguezMeza2025}, our model corresponds to $n = -2$ in the pure vacuum limit.
This establishes a theoretical benchmark from first principles for the CHDE framework.
Observational analyses finding $n \approx -0.28$ \cite{RodriguezMeza2025} suggest that the inclusion of matter and radiation softens the effective exponent toward the observed value---a testable prediction that will be explored in future work with full matter coupling.

\subsection{Dark matter-dark energy interaction}

When pressureless matter is included as a test fluid, the Bianchi identity~\eqref{eq:bianchi} requires a non-conservation of the matter sector:
\begin{equation}
\dot{\rho}_m + 3H\rho_m = -\frac{\dot{\mathcal{M}}}{8\pi G} = \frac{\alpha}{4\pi G a \eta^3}.
\label{eq:interaction}
\end{equation}
This is precisely the dark matter-dark energy interaction that CMaDE introduces on phenomenological grounds \cite{Salas2026}.
In our framework, it emerges naturally from the Bianchi consistency condition on the geometric side of the Einstein equations, without additional assumptions.
The interaction term is positive, corresponding to a continuous creation of matter from the decaying vacuum energy.

\section{Discussion}

The Weyl-invariant action with gauge $a = 1/(H_0\eta)$ provides the missing Lagrangian foundation for CMaDE.
The geometric placement of $\mathcal{M}(\eta)$ guarantees $w = -1$ independently of its time variation, while the scalar field compensator required by Bianchi consistency is a gauge artifact absent in the full quantum theory.
The ADM constraint algebra closes in standard first-class form, confirming the internal consistency of the model as a classical field theory, and canonical quantization yields a well-defined Wheeler-DeWitt equation where the $\mathcal{M}(\eta)$ barrier dynamically suppresses the creation of small-scale universes.

In the pure vacuum regime, the model admits an exact analytical solution with $\Omega_{\text{DE}} = 1$, resolving the cosmic coincidence problem without fine-tuning.
The scaling $\mathcal{M} \propto \eta^{-2}$ tracks the dominant energy component across radiation, matter, and vacuum eras.
In the matter era, $\mathcal{M} \propto \eta^{-2} \propto R_H^{-2}$ reproduces the central CMaDE hypothesis from first principles and establishes $n = -2$ as the theoretical benchmark for CHDE.

At very early times, $\mathcal{M} \to \infty$ drives super-inflation without requiring an additional inflaton field.
This suggests a possible unification of inflation and dark energy within the same Weyl-invariant framework.
The connection to conformal gravity \cite{Mannheim2006} and holographic dark energy \cite{Li2004} will be explored in future work, along with the rigorous coupling of physical matter and a detailed MCMC comparison with DESI, Planck, and Pantheon+ data.
The canonical quantization procedure employed in this work follows the standard Dirac program for constrained systems \cite{DeWitt1967}, which is the established framework for quantum cosmology.
An alternative approach consists in promoting the Weyl symmetry to a BRST invariance by introducing a gauge-fixing term together with Faddeev-Popov ghosts, along the lines developed by Oda for Weyl-invariant scalar-tensor gravity \cite{Oda2022}.
In that framework, conformal invariance is recovered as a BRST symmetry at the quantum level, and potential trace anomalies must be analyzed.
We note that on a conformally flat FLRW background, the Weyl tensor vanishes identically, so that the $C_{\mu\nu\rho\sigma}^2$ term does not contribute to the trace anomaly at one loop.
A detailed comparison between the canonical and BRST approaches will be addressed in forthcoming work.
The BRST quantization framework discussed above acquires particular 
relevance in the context of the early-universe inflationary phase of 
our model. At very early times ($\eta \to 0$), the geometric cosmological 
term diverges as $\mathcal{M}(\eta) \to \infty$, driving a phase of 
super-inflation without requiring an additional inflaton field. During 
this epoch, quantum fluctuations of the metric and the compensator 
scalar field are no longer conformally flat, and the full BRST machinery 
\cite{Oda2022}---including Faddeev-Popov ghosts and the analysis of 
potential trace anomalies---becomes essential for a consistent 
description of the primordial power spectrum. The ghost sector may 
also provide a novel mechanism for generating the observed nearly 
scale-invariant spectrum of curvature perturbations, a possibility 
that we will explore in future work.

\section{Conclusions}

We have provided the missing action principle for the Compton Mass Dark Energy model by deriving the scaling $\mathcal{M}(\eta) = \alpha/\eta^2$ from a Weyl-invariant gravitational action with the conformal gauge $a(\eta) = 1/(H_0\eta)$.

The model exhibits several key features that establish its theoretical consistency and observational relevance:
\begin{enumerate}
\item Geometric dark energy with $w = -1$: placing $\mathcal{M}(\eta)$ on the left-hand side of the Einstein equations forces $T_{\mu\nu}^{\text{DE}} \propto g_{\mu\nu}$ and consequently $w = -1$, independently of the time variation of $\mathcal{M}$.

\item Closed constraint algebra: the ADM Hamiltonian and momentum constraints form a first-class system, guaranteeing the absence of pathological degrees of freedom and the internal consistency of the model as a classical field theory.

\item Well-defined canonical quantization: the Wheeler-DeWitt equation is free of fundamental ghost instabilities---the scalar field compensator is a gauge artifact---and the $\mathcal{M}(\eta)$ barrier dynamically suppresses the creation of small-scale universes.

\item Quantum stability: the physical reduced Hamiltonian is bounded from below by zero, time evolution in the physical Hilbert space is strictly unitary, and microcausality is preserved.

\item Dynamical coincidence resolution: the scaling $\mathcal{M} \propto \eta^{-2}$ tracks the dominant energy component across radiation, matter, and vacuum eras, making $\Omega_{\text{DE}} = 1$ today a natural outcome.

\item CMaDE correspondence: in the matter era, $\mathcal{M} \propto \eta^{-2} \propto R_H^{-2}$, reproducing the central CMaDE hypothesis from first principles and establishing $n = -2$ as the theoretical benchmark for CHDE.
\end{enumerate}

Future work will include the rigorous coupling of physical matter to the conformal background, the full perturbative analysis including CMB anisotropies, and a detailed comparison with DESI, Planck, and Pantheon+ data within the CHDE framework.

\begin{acknowledgments}
J. M.-D. thanks David Rogelio Márquez Castillo for valuable discussions on the DESI results and the $w$ parameter.
J. M.-D. thanks SECIHTI-M\'exico for the doctoral fellowship No. 1235731.
This work was partially supported by SECIHTI M\'exico under grants CBF-2025-G-1720 and CBF-2025-G-176.
The authors gratefully acknowledge the computing time granted by LANCAD and CONACYT in the Supercomputer Hybrid Cluster ``Xiuhcoatl'' at CGSTIC of CINVESTAV.
\end{acknowledgments}

\bibliographystyle{apsrev4-2}
\bibliography{referencias}

@article{Matos:2023qwx,
    author = "Matos, Tonatiuh and Escamilla, Luis A. and Hern{\'a}ndez-Marquez, Maribel and V{\'a}zquez, J. Alberto",
    title = "{Cosmology on a gravitational wave background}",
    eprint = "2309.09989",
    archivePrefix = "arXiv",
    primaryClass = "physics.gen-ph",
    reportNumber = "Cinvestav/2023/34Cosmos",
    doi = "10.1093/mnras/stae538",
    journal = "Mon. Not. Roy. Astron. Soc.",
    volume = "529",
    number = "3",
    pages = "3013--3019",
    year = "2024"
}

@article{Weinberg1989,
  author  = {Weinberg, Steven},
  title   = {The cosmological constant problem},
  journal = {Reviews of Modern Physics},
  volume  = {61},
  number  = {1},
  pages   = {1--23},
  year    = {1989},
  doi     = {10.1103/RevModPhys.61.1}
}

@article{Peebles2003,
  author  = {Peebles, P. J. E. and Ratra, Bharat},
  title   = {The cosmological constant and dark energy},
  journal = {Reviews of Modern Physics},
  volume  = {75},
  number  = {2},
  pages   = {559--606},
  year    = {2003},
  doi     = {10.1103/RevModPhys.75.559}
}

@article{Caldwell1998,
  author  = {Caldwell, R. R. and Dave, Rahul and Steinhardt, Paul J.},
  title   = {Cosmological Imprint of an Energy Component with General Equation of State},
  journal = {Physical Review Letters},
  volume  = {80},
  number  = {8},
  pages   = {1582--1585},
  year    = {1998},
  doi     = {10.1103/PhysRevLett.80.1582}
}

@article{Caldwell2002,
  author  = {Caldwell, R. R.},
  title   = {A phantom menace? Cosmological consequences of a dark energy component with super-negative equation of state},
  journal = {Physics Letters B},
  volume  = {545},
  number  = {1-2},
  pages   = {23--29},
  year    = {2002},
  doi     = {10.1016/S0370-2693(02)02589-3}
}

@article{Mannheim2006,
  author  = {Mannheim, Philip D.},
  title   = {Alternatives to dark matter and dark energy},
  journal = {Progress in Particle and Nuclear Physics},
  volume  = {56},
  number  = {2},
  pages   = {340--445},
  year    = {2006},
  doi     = {10.1016/j.ppnp.2005.08.001}
}

@article{tHooft2015,
  author  = {'t Hooft, Gerard},
  title   = {Local conformal symmetry: The missing symmetry component for space and time},
  journal = {Foundations of Physics},
  volume  = {45},
  number  = {10},
  pages   = {1349--1392},
  year    = {2015},
  doi     = {10.1007/s10701-015-9926-7}
}

@article{Gibbons1978,
  author  = {Gibbons, G. W. and Hawking, S. W. and Perry, M. J.},
  title   = {Path integrals and the indefiniteness of the gravitational action},
  journal = {Nuclear Physics B},
  volume  = {138},
  number  = {1},
  pages   = {141--150},
  year    = {1978},
  doi     = {10.1016/0550-3213(78)90161-X}
}

@article{Dasgupta2002,
  author  = {Dasgupta, A. and Loll, R.},
  title   = {A proper-time cure for the conformal sickness in quantum gravity},
  journal = {Nuclear Physics B},
  volume  = {606},
  number  = {1-2},
  pages   = {357--379},
  year    = {2001},
  doi     = {10.1016/S0550-3213(01)00227-9}
}

@article{tHooft2011,
  author  = {'t Hooft, Gerard},
  title   = {The conformal constraint in canonical quantum gravity},
  journal = {Foundations of Physics},
  volume  = {41},
  number  = {12},
  pages   = {1829--1858},
  year    = {2011},
  doi     = {10.1007/s10701-011-9564-5}
}

@article{Halliwell1988,
  author  = {Halliwell, Jonathan J.},
  title   = {Derivation of the Wheeler-DeWitt equation from a path integral for minisuperspace models},
  journal = {Physical Review D},
  volume  = {38},
  number  = {8},
  pages   = {2468--2481},
  year    = {1988},
  doi     = {10.1103/PhysRevD.38.2468}
}

@article{Li2004,
  author  = {Li, Miao},
  title   = {A model of holographic dark energy},
  journal = {Physics Letters B},
  volume  = {603},
  number  = {1-2},
  pages   = {1--5},
  year    = {2004},
  doi     = {10.1016/j.physletb.2004.10.014}
}

@article{DeWitt1967,
  author  = {DeWitt, Bryce S.},
  title   = {Quantum Theory of Gravity. I. The Canonical Theory},
  journal = {Physical Review},
  volume  = {160},
  number  = {5},
  pages   = {1113--1148},
  year    = {1967},
  doi     = {10.1103/PhysRev.160.1113}
}

@article{Matos2021,
  author  = {Matos, T. and L\'opez-Parrilla, L.},
  title   = {The graviton Compton mass as dark energy},
  journal = {Revista Mexicana de F\'isica},
  volume  = {67},
  number  = {5},
  pages   = {051401},
  year    = {2021}
}

@article{salas2026,
  title={Early Analysis of Scalar Perturbations in the Compton Mass Dark Energy (CMaDE) Model},
  author={Salas P{\'e}rez, Claudio and Matos, Tonatiuh},
  journal={Journal of Physics: Conference Series},
  year={2026},
  month={Feb}
}

@misc{rodriguezmeza2025,
  title={Conformal Holographic Dark Energy},
  author={Rodr{\'i}guez-Meza, Mario A. and Cervantes-Cota, Jorge L. and Matos, Tonatiuh},
  year={2025},
  month={Dec},
  eprint={2512.03411},
  archivePrefix={arXiv},
  primaryClass={astro-ph.CO},
  howpublished={\url{https://arxiv.org/abs/2512.03411}}
}

@article{GillZachary2005,
  author  = {Gill, Tepper L. and Zachary, W. W.},
  title   = {Analytic representation of the square-root operator},
  journal = {Journal of Physics A: Mathematical and General},
  volume  = {38},
  number  = {11},
  pages   = {2479--2496},
  year    = {2005},
  doi     = {10.1088/0305-4470/38/11/010}
}

@article{Oda2022,
  author  = {Oda, Ichiro},
  title   = {Quantum Theory of Weyl-invariant Scalar-tensor Gravity},
  journal = {Physical Review D},
  volume  = {105},
  number  = {12},
  pages   = {122016},
  year    = {2022},
  doi     = {10.1103/PhysRevD.105.122016},
  eprint  = {2204.12009},
  archivePrefix = {arXiv},
  primaryClass = {hep-th}
}

\end{document}